\documentclass[twocolumn]{webofc}

\usepackage[varg]{txfonts}   % Web of Conferences font
\usepackage{hyperref}
\usepackage{url}
\hypersetup{colorlinks=true,citecolor=blue,urlcolor=blue,linkcolor=blue}
\begin{document}
\title{Clusterization aspects in the states of non-rotating and  fast rotating nuclei}
%
% subtitle is optional

\author{\firstname{A.V.} \lastname{Afanasjev}\inst{1}\fnsep\thanks{\email{Anatoli.Afanasjev@gmail.com}} 
% \ and
%        \firstname{Isabelle} \lastname{Houlbert}\inst{2}\fnsep\thanks{\email{Mail address for second
%             author if necessary}} 
%\and
%        \firstname{Agnès} \lastname{Henri}\inst{3}\fnsep\thanks{\email{Mail address for last
%             author if necessary}}
        % etc.
}

\institute{Department of Physics and Astronomy, Mississippi State University, MS 39762
%\and
%           the second here 
%\and
%           Last address
          }

\abstract{The understanding of clustering aspects at the ground state of nuclei and in 
fast rotating ones within the framework of covariant density functional theory has been 
reviewed and reanalyzed. The appearance of many exotic nuclear shapes in nuclear 
chart can be inferred from combined analysis of nodal structure of the densities of the 
single-particle states and the evolution of such states in the Nilsson diagram with 
deformation and particle number.  Such analysis which is supported by fully self-consistent 
calculations allows to predict the existence of nuclear configurations with specific shape 
or cluster properties at ground state and at high spin.  For example, it indicates  that  in 
a given shell with principal quantum number $N$ only the lowest in energy two-fold degenerate  
deformed state can contribute to the formation of linear chain $\alpha$ cluster structures.
}
\maketitle
%
%%%%%%%%%%%%%%%%%%%%%%%%%%%
\section{Introduction}
\label{intro}
%%%%%%%%%%%%%%%%%%%%%%%%%%%

   For a long time  light nuclei have been a playground for nuclear physicists
to study the existence and coexistence of different types of nuclear shapes
such as  $\alpha$-clustered, nuclear molecules, the combination of large
spheroidal-like cluster and $\alpha$-particles, ellipsoidal like ones etc
(see, for example,  Refs.\ \cite{EKNV.12,FHHKLM.18}). Among recent 
interesting results is the experimental observation of rotational band in 
$^{14}$C which is built on linear chain of three $\alpha$-particles plus two
neutrons (see Ref.\ \cite{C14-lin-chain.23}) and theoretical investigation
of the $\alpha$ clusterization in the $^{8}$Be and $^{12}$C nuclei in 
ab-initio calculations (see Ref.\ \cite{Otsuka-clust.22}).

Above mentioned shapes have been analysed in the framework of different 
theoretical models.  One class of such models is the cluster models.  It provides 
an important insight into cluster dynamics of nucleus but  is not able to describe 
many shell model configurations and is limited by model assumptions. Alternatively, 
the formation of the clusters and other exotic shapes via microscopic single-nucleon
degrees of freedom and many-body correlations can be described in density functional 
theory (DFT) (see Refs.\ 
\cite{ER.04,ASPG.05,RMUO.11,EKNV.12,EKNV.14,YIM.14,RA.16,AA.18,ATT.24}).

 In the present paper, I review and reanalyze the results obtained in the framework
of cranked relativistic mean field theory in Refs.\ \cite{RA.16,AA.18,ATT.24} and present
some new results.  Note that in this paper we restrict our consideration to reflection-symmetric 
shapes characterized by extreme deformation. A particular attention 
is paid to the correlations between nodal structure of the density distributions  of the 
single-particle states and the creation of  exotic  shapes as a function of particle number 
and deformations. The impact of collective rotation on these features is also analyzed.

%%%%%%%%%%%%%%%%%%%%%%%%%%%%%%
\section{Nodal structure of the wavefunctions of occupied 
             single-particle states:  a key to understanding the 
             clusterization and nuclear molecules in finite nuclei}
\label{nodal-structure}
%%%%%%%%%%%%%%%%%%%%%%%%%%%%%%

\begin{figure*}[htb]
\centering
\includegraphics[width=18.2cm,clip]{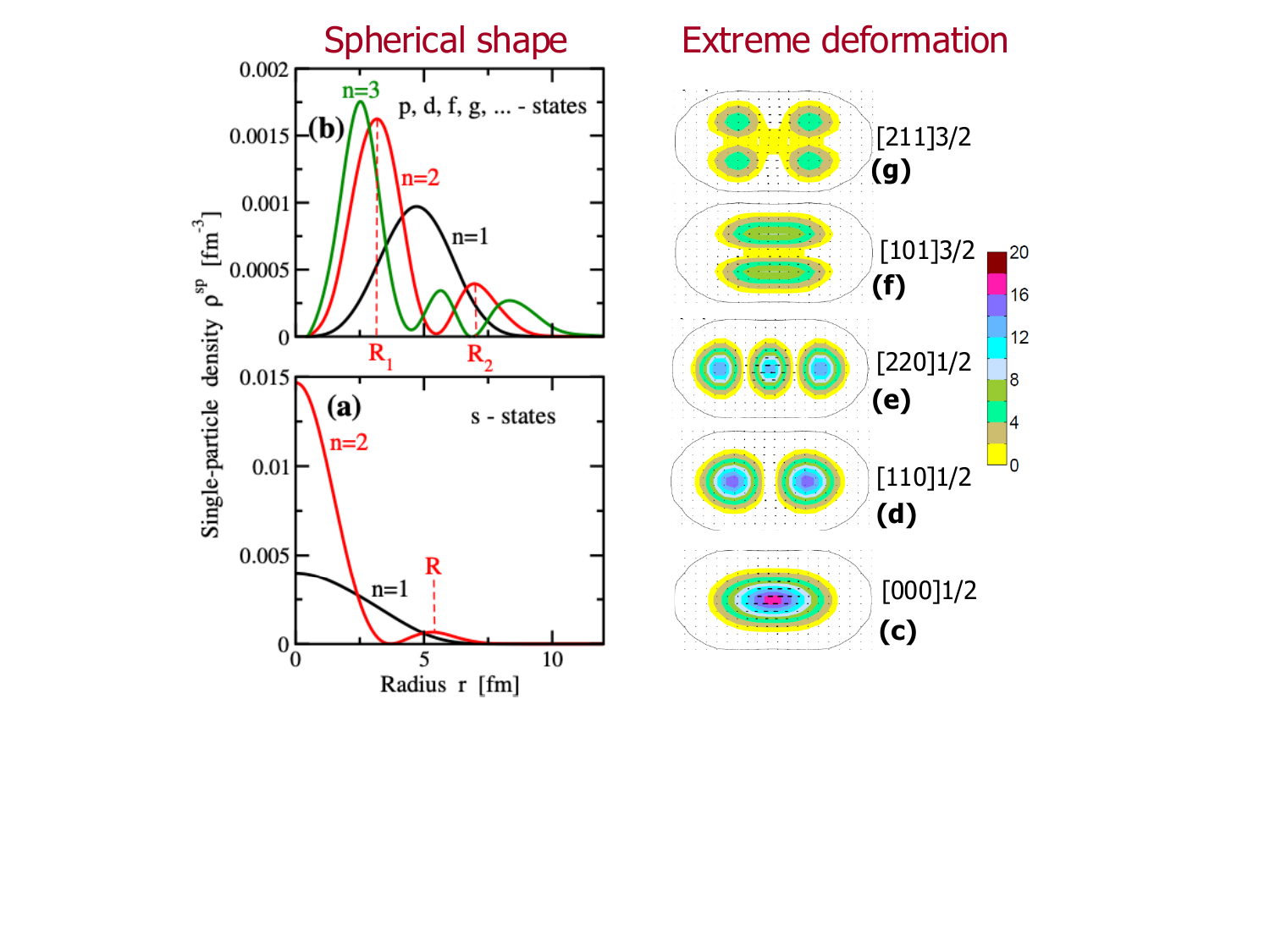}
\vspace{-4.0cm}
\caption{Left panels: schematic illustration of the nodal structure of  
density for different single-particle states at spherical shape (see Refs.\ 
\cite{PA.22,PA.23} for more detailed information). Right panels: the pattern 
of nodal structure of the density of the single-particle states at extreme 
deformation (for analysis of three-dimensional density distributions see
Ref.\ \cite{AA.18}). The neutron densities are plotted for indicated Nilsson states 
in  $^{40}$Ca at deformation $\beta_2\approx 1.6$ which roughly corresponds  
to megadeformed configurations in this nucleus.  The shape of the nucleus is 
shown by black solid line.  The colormap shows densities as multiplies of 0.001 
fm$^{-3}$.  The plotting of the densities starts from  0.001 fm$^{-3}$.
\label{fig-nodal}
}
\end{figure*}

  The phenomenon of clusterization in finite nuclei is intimately
connected with the nodal structure of the wavefunctions and
densities of  occupied single-particle
states. To better understand its role let me start from the analysis of 
the single-particle densities at spherical shape. The single-particle 
densities of the $s$ states have a maximum at radial coordinate 
$r=0$ (see Fig.\ \ref{fig-nodal}(a)). This is because their orbital 
angular momentum is $l=0$ and thus there is no centrifugal 
barrier. Dependent on principal quantum number $n$ their
densities look like a sphere of variable density with maximum
density at center (see the $n=1$ curve in Fig.\ \ref{fig-nodal}(a))
or as a combination of similar sphere and a spherical 
shell(s)\footnote{In geometry, a spherical shell is a region between 
two concentric spheres of differing radii.} of
variable density: the maximum of the density of the shell is located 
at $R$ (see the $n=2$ curve in Fig.\ \ref{fig-nodal}(a)).
 The measurements of the single-particle densities of the $3s$
state in spherical nuclei near $^{208}$Pb confirm this picture 
(see Ref.\ \cite{3s-density.PRL.82,3s-density-NPA.83}).

In contrast no density can be built at $r=0$ for the single-particle states which 
have  orbital momentum $l\neq 0$ (i.e. the $p$\footnote{Note 
that in relativistic framework the $p$ states have some part of small 
component of the Dirac spinor in the $s$ states which build the 
density at $r=0$, see Sec. IV of Ref.\ \cite{PA.22}.}, $d$, $f$, $g$ 
... single-particle states) due to  the presence of centrifugal barrier.
The  densities of the single-particle states can be viewed as a combination 
of $n$ spherical shells the maxima of the density of which appear
at the $R_i$ radial coordinates (see Fig.\ \ref{fig-nodal}(b)). In both
cases the number of independent structures (sphere and/or spherical
shells) is equal to principal quantum number $n$ and the node(s) of 
the single-particle wave functions at which $\rho=0$ define the natural 
boundary between these these structures (see Ref.\ \cite{PA.23}). Note 
that the nodal structure of the wavefunctions and densities of spherical 
single-particle states plays very important role in our understanding  of 
the formation of bubble structures in the nuclei  (see Ref.\ \cite{PA.22}) 
and the buildup of differential charge radii as a function of neutron number 
in isotopic chains (see Ref.\ \cite{PA.23}).

%%%%%%%%%%%%%%%%%%%%%%%%%%%%%%%%
%%%%%%%%%%%%%%%%%%%%%%%%%%%%%%%%

   The situation is drastically different for extremely deformed structures 
such as hyper- and megadeformed ones. In some instances such structures are 
denoted as rod-shaped (see Refs.\ \cite{ZIM.15,AA.18}). For such shapes, the single-particle 
states are characterized by asymptotic quantum numbers $[Nn_z\Lambda]\Omega$ 
(Nilsson quantum numbers) of the dominant component of the wave function. 
Their wave functions $\Psi_{[Nn_z\Lambda]\Omega}$ are  expanded into the 
basis states $|N'n'_z\Lambda'\Omega'>$ by
\begin{eqnarray}
\Psi_{[Nn_z\Lambda]\Omega} = \sum_{N',n'_z,\Lambda',\Omega'} c_{N'n'_z\Lambda'\Omega'} |N'n'_z\Lambda'\Omega'>.
\label{wave-funct}
\end{eqnarray}
Here, the basis states are characterized by principal quantum number $N'$,
the number $n'_z$ of nodes in the axial direction ($z$-direction)  and the 
projections of orbital ($\Lambda'$) and total ($\Omega'$) single-particle angular 
momenta on the axis of symmetry. The sum in Eq.\ (\ref{wave-funct}) runs 
over all allowable combinations of the quantum numbers $N', n'_z, \Lambda'$ 
and $\Omega'$.

   In the context of the discussion of clusterization it is important that  the 
single-particle wavefunctions at extremely large deformation are  dominated by 
a single basis state. This is the case for the configurations of interest of light 
nuclei (see Ref.\ \cite{AA.18}).  This basis state defines the dominant nodal structure of 
the wave function of the single-particle state and its spatial density distribution.  
It turns out that  only two types of the states i.e. $|N,n_z,\Lambda> =|N,N,0>$\footnote{
The nodal structure of the single-particle wave function and thus of its density 
distribution is defined by the spatial part of the wave function. Thus, at no rotation it is 
sufficient to specify only $\Lambda$ and to drop $\Omega$.} 
 and $|N,n_z,\Lambda> =|N,N-1,1>$ are important for the nuclei with masses 
$A \leq 50$ (see Ref.\ \cite{AA.18}). The density distribution of the basis states with the 
$|NN0>$ structure is axially symmetric.  It consist of the $N+1$ clusters the maxima 
of the density of  which are located at the axis of symmetry.  In contrast, the density 
distribution of  the basis states with  the $|N,N-1,1>$ structure is build of the $N$ 
donuts (or ring-like structures) the axis of the symmetry of which coincides with  the axis 
of symmetry of  the nucleus. Note that the nodal structure of the density distribution of the basis
 state is defined by the condition $N' = n_z' +2n'_r +|\Lambda'|$ where $n'_r$
 is the number of radial nodes. In the light  nuclei, the basis states of interest have 
 $n'_r=0$, so above mentioned condition is simplified to $N' = n_z'  +|\Lambda'|$.
 
 %%%%%%%%%%%%%%%%%%%%%%%%%%%%%%%%
\begin{figure}[ht]
\centering
%\vspace*{1cm}       % Give the correct figure height in cm
\includegraphics[width=8cm,clip]{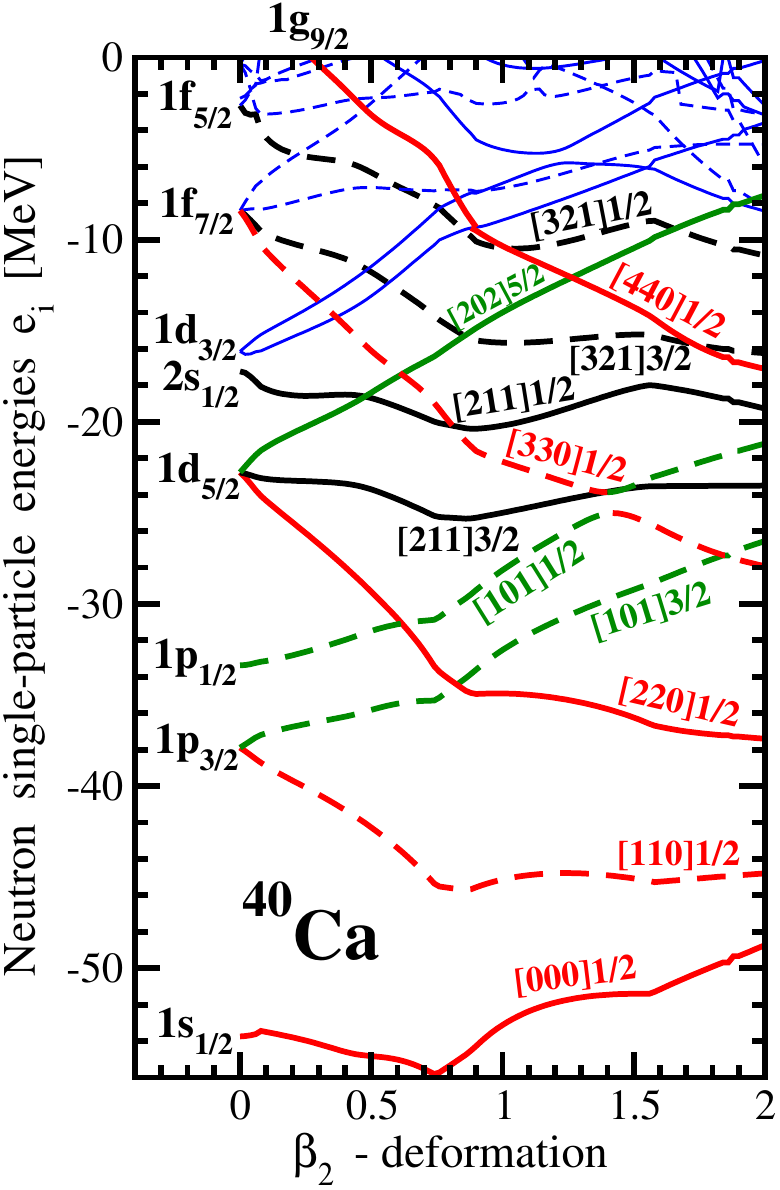}
\caption{The energies of the single-particle states in $^{40}$Ca as a 
function of quadrupole deformation $\beta_2$. They are calculated 
along the deformation path of  the lowest in total energy solution in axial relativistic Hartree-Bogoliubov 
framework \cite{AARR.14} using the NL3* CEDF \cite{NL3*}.  The orbitals are 
labelled by means of asymptotic quantum numbers  (Nilsson labels) 
$[Nn_z\Lambda]\Omega$.  Solid and dashed lines are used for
positive and negative parity states, respectively.  Note that in the
case of the $[330]1/2$ and $[440]1/2$ single-particle states, the structure
of these states are traced before and after the crossing of these
states with another ones. See text for further details.
\label{fig-Nilsson}
}
\end{figure}
%%%%%%%%%%%%%%%%%%%%%%%%%%%%%%%%
 
    Fig.\ \ref{fig-nodal} illustrates this situation. Let me first consider the states
with the $|NN0>$ structure at extreme deformation. The $[0,0,0]1/2$ deformed state 
emerges from the spherical $1s_{1/2}$ state (see Fig.\ \ref{fig-Nilsson}) which has orbital
angular momentum $l=0$. Because of the latter fact extreme 
deformation $\beta_2\sim 1.6$ leads only to the elongation of its single-particle 
density [see Fig.\ \ref{fig-nodal}(c)].  The lowest in energy extremely  deformed state 
of the $N=1$ shell (i.e. the [110]1/2 state)  emerges from the spherical $1p_{3/2}$ 
state (see Fig.\ \ref{fig-Nilsson}) which has  $l=1$.  Its single-particle density is formed 
of two clusters separated  by the gap [see Fig.\ \ref{fig-nodal}(e)].   The $[220]1/2$ 
extremely deformed state emerges from the lowest state of the $N=2$ shell (i.e. the 
$1d_{3/2}$  one):  its density consist of three clusters  [see Fig.\ \ref{fig-nodal}(d)].

   The nodal structure of the density distribution of single-particle states in spherical 
nuclei is defined by the principal quantum number $n$. In contrast, it is defined by the 
$n_z$ value of the Nilsson label of the single-particle state in extremely deformed nuclei: 
the number of density clusters along the symmetry axis is equal to $n_z+1$ in the single-particle
states with the $|NN0>$ structure.  That means that the memory of the nodal structure of
the density distribution of single-particle state at spherical shape is completely lost on 
transition to extremely deformed shapes. Note that this result is also confirmed by the 
analysis of the $[330]1/2$, $[440]1/2$ and $[550]1/2$ single-particle states in the light 
nuclei (see Figs. 4 and 9 in Ref.\ \cite{AA.18} and Fig. 8 in supplemental material to Ref.\ 
\cite{AA.18}) and in the $[770]1/2$ state of yrast superdeformed band of $^{152}$Dy (see Fig. 
9(a) in Ref.\ \cite{TO-rot}).

    Doughnut-like structures of the single-particle density distribution at extreme 
deformation is illustrated by the example of the $[101]3/2$ and $[211]3/2$ 
states [see Figs.\ \ref{fig-nodal}(f) and(g)]. The first state emerges from the
$1p_{3/2}$ spherical subshell (see Fig.\ \ref{fig-Nilsson}) and its single-particle 
density has single dooughnut structure.  The second one originates from the $2d_{3/2}$ 
spherical subshell  (see Fig.\ \ref{fig-Nilsson}): its single-particle density looks 
like  two doughnuts located symmetrically with respect of the center of the nucleus. 
Higher number of donuts in the single-particle density is also possible (see,
for example, Fig.\ 8 in Ref.\ \cite{AA.18}).

The wave functions of normal deformed nuclei with quadrupole deformation  
$\beta_2 \approx 0.3$  are dominated by few spherical components. This is 
because the shapes of such nuclei do not deviate substantially from 
spherical ones.  Thus, the densities of the single-particle states at such 
deformations are the mixtures of the densities shown in left panels of 
Fig.\ \ref{fig-nodal} and the formation of well separated clusters or 
donuts seen at extreme deformation does not take place at normal
deformation.  As a result, one can conclude that the clusterization features
do not appear in the single-particle structures at spherical and normal 
deformed shapes in finite nuclei.

     The Nilsson diagram for the states of interest is shown in Fig.\ 
\ref{fig-Nilsson}. The single-particle states for which detailed analysis of the nodal 
structure of the density distribution has been carried out in the present paper and in 
Ref.\ \cite{AA.18} are shown by thick lines of different color.  The states for which no 
such analysis is available are shown by thin blue lines but they do not play a role in 
the configurations of interest in the $A\leq 50$ nuclei. At extreme deformation, all the 
states shown by thick lines can be separated into three classes:

\begin{itemize}

\item
 {\it cluster forming states} i.e. the single-particle states dominated by the 
$|N,N,0>$ basis state which leads to a formation of the $N+1$ spheroidal-like 
density clusters along the axis of symmetry of the nucleus. These states are 
shown by red thick lines in Fig.\ \ref{fig-Nilsson}.

\item
  {\it single doughnut states} i.e. the single-particle states which are dominated 
by the $|N,0,1>$ basis state the density distribution of which is represented 
by a single doughnut.  They are shown by green thick lines in 
Fig.\ \ref{fig-Nilsson}.

\item
  {\it multiply doughnut states} i.e. the single-particle states the density of 
which is defined by the $|N,N-1,1>$ $(N\geq 2)$ basis state which 
contains  $N$ doughnuts. They are shown by black thick lines in Fig.\ 
\ref{fig-Nilsson}.

\end{itemize}

       The present analysis of the nodal structure of the single-particle densities
at extreme deformation indicates that  linear $\alpha$-chain structures (such 
as those shown in Fig. 1 of Ref.\ \cite{OFE.06} and Fig. 4 of Ref.\ \cite{EKNV.14})  
can only be formed by the occupation of  cluster forming states. These are $2\alpha$, 
$3\alpha$,  $4\alpha$, and  $5\alpha$ linear-chain structures in $^{8}$Be, 
$^{12}$C, $^{16}$O and $^{20}$Ne nuclei, respectively. The present analysis for 
the first time reveals that there is only one two-fold degenerate deformed single-particle state in a given
$N$-shell which acts as $\alpha$-cluster forming state.  These are the lowest 
in energy states in a given $N$-shell  shown by red thick lines in Fig.\ 
\ref{fig-Nilsson}.  Note that above mentioned multiply $\alpha$-cluster structures 
are located at very high excitation energies with respect of the ground states
of the nuclei under discussion.

  The occupation of single and/or multiply dooughnut states will lead to suppression 
of the $\alpha$-clusterization and formation of ellipsoidal and nuclear molecule
shapes. The analysis of the nodal structure of the single-particle densities at
extreme deformation  (see Fig.\ \ref{fig-nodal}) and Nilsson diagram (see Fig.\ 
\ref{fig-Nilsson}) leads to conclusion that only for a very few specific nucleonic 
configurations the processes of formation of the $\alpha$-cluster structures is possible.  
There  are several reasons for that.  First, the portion of cluster forming single-particle 
states in a given $N$ shell drastically decreases with increasing principal quantum 
number $N$: such states represent only 100\%, 33\%, 15\%, 10\%, and 6.5\% of all 
single-particle  states in the $N=0$, 1, 2, 3 and 4 shells, respectively. Second, because
of the rarity of cluster forming single-particle states this  process is possible only in
relatively light nuclei since with increasing proton and neutron numbers the
process of the occupation of single and/or multiply doughnut states becomes inevitable
(see Fig.\ \ref{fig-Nilsson}). Third, with increasing particle number the contribution of the 
states favoring $\alpha$-clusterization into the structure of total wavefunction decreases 
and  ellipsoidal-like shapes  become dominant species of nuclear shapes. 

   Considering  typical pattern of density distributions of single-particle states 
presented in right panels of  Fig.\ \ref{fig-nodal} and the arrangement of the
single-particle states at extreme deformation (see Fig.\ \ref{fig-Nilsson}) one can easily
predict the existence of exotic structures which are built from large central spheroidal-like
structure and two alpha  particles on opposite sides of it located at the same axis of symmetry.
The examples of density distributions of such exotic  structures are shown in Fig. 4 
of Ref.\ \cite{EKNV.14} (second from bottom density distribution in the $^{20}$Ne 
nucleus and  third one in $^{24}$Mg).  For example, the configuration $confA(^{16}{\rm O})=[000]1/2 \otimes
[110]1/2 \otimes [220]1/2 \otimes [101]3/2$\footnote{For simplicity of discussion we consider
here only the $N=Z$ nuclei with $A=4Z$. Thus, each of single-particle states in the 
configuration label is occupied by two protons and two neutrons.} in $^{16}$O 
is an example of such exotic structure.
The occupation of first three orbitals by protons and neutrons leads to triple 
linear $\alpha$ chain similar to that which exists in $^{12}$C (see Fig.\ 1(a) in Ref.\ 
\cite{AA.18} and Fig. 2 in Ref.\ \cite{ZIM.15}).  The $[101]3/2$ orbital has single doughnut
structure of single-particle densities
[see Fig.\ \ref{fig-nodal}(f)]: its occupation in combination with central $\alpha$ cluster 
leads to the formation of central spheroidal-like structure.  Note that Fig.\ \ref{fig-Nilsson}
suggests that this configuration is the lowest in energy at extreme deformation
$\beta_2 = 1.2 \pm 0.4$.

  Similar exotic structures can be built starting from  $confA(^{16}$O) in heavier nuclei. 
For example, the addition of two protons and two neutrons leads to the configurations 
$confA(^{20}{\rm Ne}) = confA(^{16}{\rm O}) + [101]1/2$ and 
$confB(^{20}{\rm Ne})=confA(^{16}{\rm O}) + [303]1/2$ in $^{20}$Ne (see Fig.\ \ref{fig-Nilsson}) 
which have similar exotic structure. Note that central spheroidal-like structure will be more 
pronounced in the former configuration  since the density of the $[101]1/2$ state has single 
doughnut structure (see Fig.\  \ref{fig-nodal}(f)).  Note that the latter configuration is built from 
linear $4\alpha$ chain, which is analyzed in the case of $^{16}$O in Refs.\ \cite{IMIO.11,YIM.14},
and the doughnut density distribution generated by the $[101]3/2$ state.  Nilsson diagram of Fig.\ 
\ref{fig-nodal} suggests that  $confA(^{20}{\rm Ne})$ is more energetically favored as compared 
with $confB(^{20}{\rm Ne})$ at $\beta_2 \leq 1.3$ but the situation becomes reversed at 
higher deformations.  Fig.\ \ref{fig-Nilsson} suggests that  the
$confA(^{24}{\rm Mg})=[000]1/2 \otimes [110]1/2 \otimes [220]1/2 \otimes  [330]1/2 \otimes [101]3/2 \otimes [101]1/2$
configuration in $^{24}$Mg should be close to the lowest one at $\beta_2 \geq 1.2$. The occupation
of the first four orbitals builds linear $4\alpha$ chain while the occupation of the last two ones builds
single doughnut structure around an axis of symmetry of the nucleus located at its center. Resulting total
density is close to the one shown in Fig. 4  of Ref.\ \cite{EKNV.14} (third from bottom density distribution
in $^{24}$Mg). Note that this configuration can be rewritten as
$confA(^{24}{\rm Mg})=confA(^{16}{\rm O}) \otimes [330]1/2 \otimes [101]1/2$.

    The formation of  ellipsoidal-like shapes at extreme deformation requires the occupation 
of all three classes of single-particle states, i.e.  cluster forming states, which built density 
along the axis of symmetry  of the nucleus, and single and  multiply doughnut states.  This process 
starts when the $[211]3/2$, $[211]1/2$, $[321]3/2$ and $[321]1/2$ states which have multiply 
doughnut structure of the single-particle density and which are shown by thick black
lines in Fig.\ \ref{fig-Nilsson} start to be occupied in the configurations of interest. 
To build the nuclear molecules  one has to move the matter from the neck (equatorial) region 
into the polar regions of the nucleus. To achieve that one should remove the particles  from the 
states which have single doughnut structure of the single-particle density (i.e. single-particle states 
with the $|N,0,1>$ structure shown by thick green lines in Fig.\ \ref{fig-Nilsson}). Some
examples of particle-holes excitations leading to nuclear molecules have been discussed
in Ref.\ \cite{AA.18}.

%%%%%%%%%%%%%%%%%%%%%%%%%%%%%%%%
\section{The role of rotation in the clusterization of atomic nuclei}
\label{rotation}
%%%%%%%%%%%%%%%%%%%%%%%%%%%%%%%%

%%%%%%%%%%%%%%%%%%%%%%%%%%%%%%%%
\begin{figure}[ht]
\centering
%\vspace*{1cm}       % Give the correct figure height in cm
\includegraphics[width=10.0cm,clip]{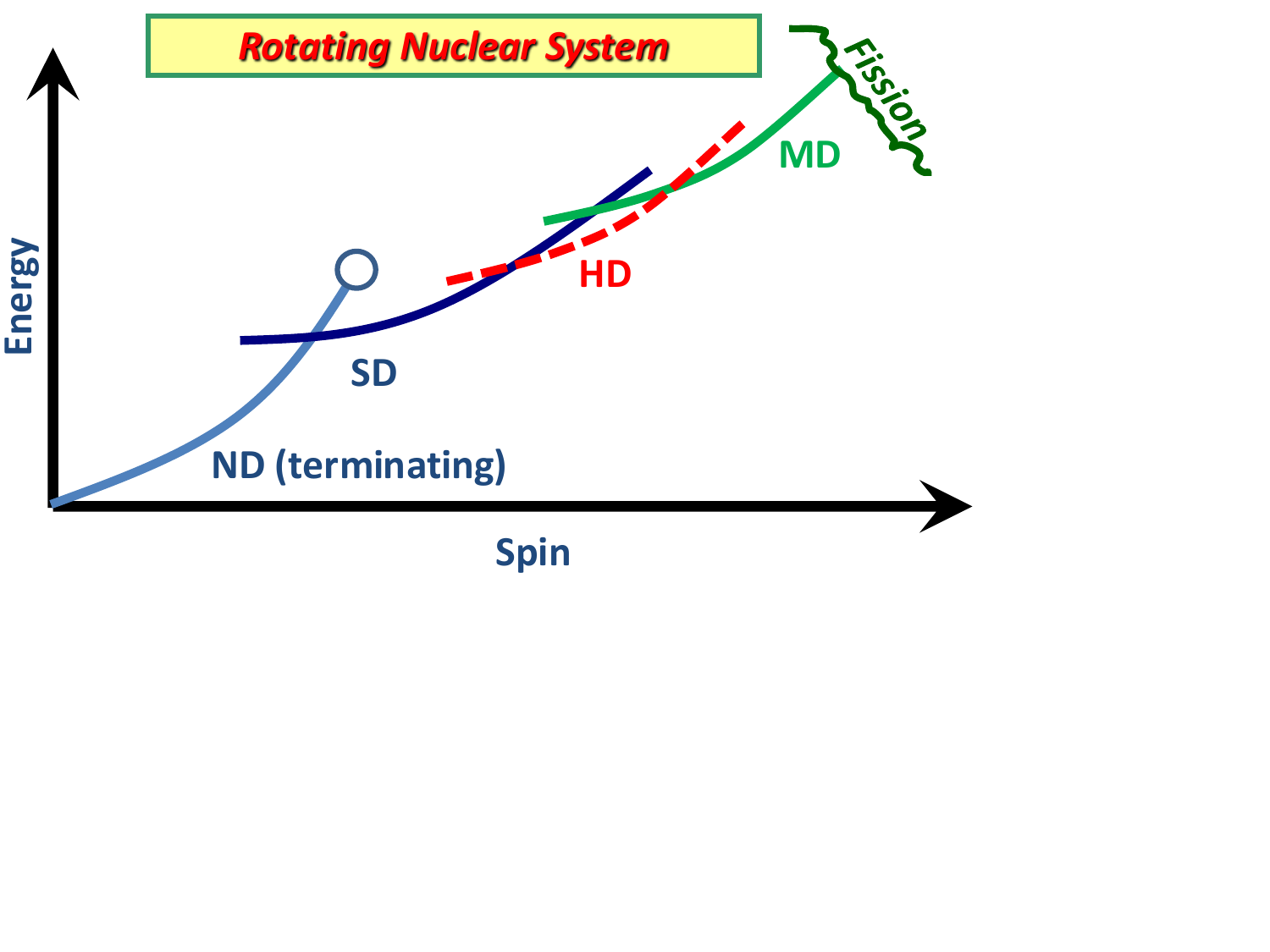}
\vspace{-3.5cm}
\caption{General structure of rotational spectra in light $(A<50)$ nuclei.
Based on the results of systematic investigations presented in Refs.\ 
\cite{RA.16,ATT.24}.
%ND, SD, HD and MD are abbreviations for normal-, super-, 
%hyper- and megadeformed rotational bands. 
Terminating state of normal deformed
band is shown by open circle. See text for further details.}
\label{fig-rotation}
\end{figure}
%%%%%%%%%%%%%%%%%%%%%%%%%%%%%%%%

     Unfortunately, at spin zero the absolute majority of predicted extremely
deformed structures cannot be observed in experiment. This is because 
they are located at very high excitation energies of 15 to 40 MeV and in 
many cases no local minimum corresponding to such shapes exists in 
potential energy surfaces (see, for example, Fig. 2 in Ref.\ \cite{EKNV.14} and 
Fig.\ 1 in Ref.\ \cite{YIM.14}). The collective rotation of the nuclei makes the 
conditions  for the observation of extremely deformed structures at high spin
more favorable than at low one because of the following reasons:

\begin{itemize} 
\item
    The angular momentum of the rotational bands build on the ground state and 
low excited configurations in the nuclei of interest is quite limited. For example, 
ground state rotational band of $^{20}$Ne terminates at $I=8^+$ and in some
of the nuclei of interest such termination is expected even at lower spins. Thus, 
above such spins the yrast line is predicted to be formed by the configurations which  
are built by sequential particle-hole excitations leading to larger angular momentum 
content and larger deformations of these configurations (see Refs.\ \cite{RA.16,ATT.24}).
This is schematically illustrated in Fig.\ \ref{fig-rotation} where along the yrast line with 
increasing spin normal deformed (ND) configurations are first replaced by superdeformed 
(SD) ones,  then by hyperdeformed (HD) ones and finally by megadeformed (MD) 
configurations. The experimental spectra of $^{40}$Ca confirms this structure and
shows the transition from spherical ground state at $I=0$ to the ND structures at moderate
spins and then to the SD rotational band at high spins (see Ref.\ \cite{Ca40-PRL.01}).
 
\item
   The rotation leads to the change of the energies of the single-particle states and
the emergence of new shell closures (see, for example, Figs. 2-4 in Ref.\ \cite{RA.16})
which stabilizes respective extremely deformed minima at rotation. Especially important 
role in this process is played by intruder orbitals emerging from the lowest deformed 
state of a given $N$ shell (see above mentioned figures). As discussed in Sec.\ 
\ref{nodal-structure}, these are cluster forming states which play important role in the
formation of linear $\alpha$ cluster structures in light nuclei.

\item
  At the spins of interest static pairing correlations in extremely deformed configurations 
are substantially quenched by a combination of factors which include Coriolis antipairing 
effect induced by collective rotation \cite{RS.80,VDS.83},  the  emergence of large shell 
gaps \cite{SGBGV.89} and blocking effect induced by particle-hole excitations \cite{RS.80}.
This leads to  a significant reduction or even suppression of  configuration mixing induced 
by pairing correlations.  As a consequence, the structure of total density of nucleonic 
configurations is defined by the single-particle densities of occupied states and their
nodal structure discussed  in Sec.\  \ref{nodal-structure}. In addition, weak or no pairing is 
expected to lead to a substantial increase of the fission barrier as compared with the case 
of regular pairing (see Ref.\ \cite{KALR.10}). This will extend the spin range at which 
rotational states can exist to a higher limit. 
    
\end{itemize}

   On the one hand, the collective rotation leads (via Coriolis interaction) to some reduction 
of the weight of the dominant basis state in the single-particle wave function and to some 
delocalization of single-particle density (i.e. to some suppression of both  $\alpha$-clusterization
and the formation of nuclear molecules)  (see Ref.\ \cite{AA.18}). However, for the single-particle 
states located at  and near the bottom of nucleonic potential  (i.e. for those shown by thick lines 
in Fig.\  \ref{fig-Nilsson}) the rotation preserves the nodal structure of the densities of the 
single-particle states (see Ref.\ \cite{AA.18}). On the other hand, collective rotation 
suppresses pairing interaction and as a consequence the configuration mixing induced by 
it and brings extremely deformed rotational states to the yrast line. The former feature counteracts 
some suppression of $\alpha$-clusterization and the formation of nuclear molecules induced 
by rotation at the level of the structure of single-particle states. The latter one favors
the population of such states in ion collisions which potentially makes experimental observation
of extremely deformed states at high spin  feasible.

%%%%%%%%%%%%%%%%%%%%%%%%%%%%%%%%
\section{New type of cluster structures}
\label{rotation}
%%%%%%%%%%%%%%%%%%%%%%%%%%%%%%%%
 
 %%%%%%%%%%%%%%%%%%%%%%%%%%%%%%%%
\begin{figure}[ht]
\centering
%\vspace*{1cm}       % Give the correct figure height in cm
\includegraphics[width=12.3cm,clip]{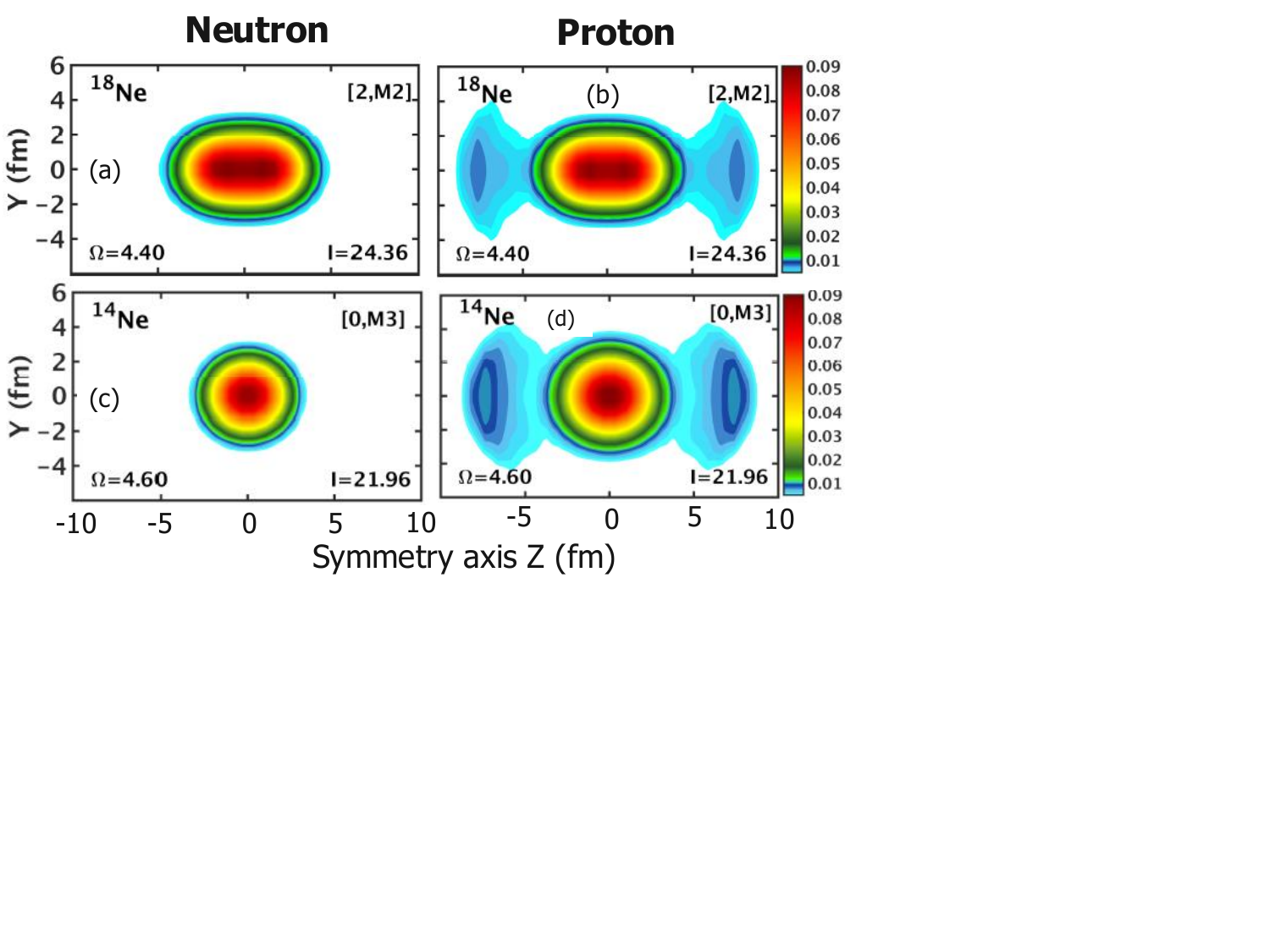}
\vspace{-4.3cm}
\caption{Neutron (left column) and proton (right column) density distributions 
of the configurations in $^{18}$Ne and $^{14}$Ne at indicated spins and 
frequencies.  The density colormap starts at $\rho=0.005$ fm$^{-3}$ and 
shows the densities in fm$^{-3}$. The results for $^{14}$Ne are taken from
Ref.\ \cite{ATT.24}. For details of configuration labelling see Sec. 5.2.2 in
Ref.\ \cite{ATT.24}. }
\label{fig-cluster-densities}
\end{figure}
%%%%%%%%%%%%%%%%%%%%%%%%%%%%%%%%

    As discussed above, the lowest in energy deformed states emerging 
from each $N$ shell have the $[NN0]1/2$ structure. These states are
the only ones which lead to the $\alpha$ clusterization. One can see that
with increasing deformation such states shown by thick red lines in 
Fig.\ \ref{fig-Nilsson} come close to each other. This explains why
linear chain $n\alpha$ cluster configurations such as $[000]1/2 \otimes [110]1/2$
in $^{8}$Be,  $[000]1/2 \otimes [110]1/2 \otimes [220]1/2$ 
in $^{12}$C and $[000]1/2 \otimes [110]1/2 \otimes [220]1/2 \otimes [330]1/2$ 
in $^{16}$O appear as the lowest in energy at extreme deformations.

    There are two peculiar features about the $[NN0]1/2$ single-particle states.
First, in the single-particle densities the maxima of the density of the clusters 
located farthest away (in polar region)  from the center of nucleus are substantially higher than those 
for the clusters located at or near the center of the nucleus.  This feature is 
especially pronounced for  $N\geq 3$ (see Figs. 4 and 9 in Ref.\ \cite{AA.18}, 
Fig. 8 in supplemental material to Ref.\ \cite{AA.18}) and Fig. 9(a) in Ref.\ 
\cite{TO-rot}).  Second, the lowest in energy signature branch of such states
has the largest single-particle angular momentum alignment in a given $N$
shell. Because of this feature such states are strongly down-sloping as a function 
of rotational frequency  $\Omega_x$ (see, for example, the evolution of the 
$[440]1/2(r=-i)$ orbital in Figs. 2, 3 and 4 in Ref.\ \cite{RA.16}).   As a result, 
nucleonic configurations based on the occupation of such hyper- and mega-intruder 
orbitals become yrast or near yrast at high angular momentum in the calculations. 
Two examples of such configurations are discussed below.

   Note that these orbitals emerge two or three shells above the normal shell
in which the Fermi level is located at normal deformation.  As a consequence, 
these orbitals have substantially larger radius of the single-particle states than
those located in the normal shell (see, for example, Table 1 in Ref.\ \cite{PA.23}).
Moreover, because of extreme deformation there is a very strong  $\Delta N=2$ 
mixing in the structure of the single-particle wave functions (see Sec. 5.2.2 in 
Ref.\ \cite{ATT.24}). The combination of these
factors with the concentration of the matter of the $[NN0]1/2$ states in polar 
regions of nucleus leads to cluster-type giant proton halo in proton subsystem 
of rotational bands $^{14,18}$Ne (see Fig.\ \ref{fig-cluster-densities}). It  consist 
of high density  prolate central cluster and low density ($\rho_\pi \approx 0.01$ 
fm$^{-3}$) clusters centered at around $z \approx  \pm 7$ fm. Note that $^{18}$Ne 
is the last proton-bound nucleus at spin zero. In  contrast,  the rotational band 
in $^{14}$Ne becomes proton bound at high spin (see Ref.\ \cite{ATT.24}). 

%%%%%%%%%%%%%%%%%%%%%%%%%%%%%%%%
\section{Conclusions}
\label{concl}
%%%%%%%%%%%%%%%%%%%%%%%%%%%%%%%%

   In summary, the most important features of the nodal structure 
of the densities of the single-particle states at extreme deformation
and their impact on $\alpha$-clusterization, the formation of exotic
shapes, nuclear molecules and ellipsoidal shapes have been 
analyzed. It turns out that the appearance of many exotic shapes 
in nuclear chart as a function of particle number can be inferred
from combined analysis of typical pattern of single-particle densities
and the evolution of the single-particle states in the Nilsson diagram
with deformation and particle number. Such analysis allows to predict 
the existence of nuclear configurations with specific shape or cluster properties. 
For example, it indicates  that only the lowest in energy two-fold degenerate 
deformed state of a given $N$-shell can contribute to the formation of the linear
chain $\alpha$ cluster structures.  Although such an analysis may look 
schematic, its validity is confirmed by a detailed fully self-consistent 
cranked relativistic mean field calculations presented in Refs.\ 
\cite{RA.16,AA.18,ATT.24}.  Note that collective rotation acts as a tool
to bring exotic shapes of interest from high excitation energies at zero
spin down to the yrast line at high spin.

\bibliography{references-44-next-gen-CEDFs-dub.bib}

\begin{thebibliography}{27}

\bibitem{EKNV.12}
J.P. Ebran, E.~Khan, T.~Nik\v{s}i\'{c}, D.~Vretenar, Nature \textbf{487}, 341
  (2012).

\bibitem{FHHKLM.18}
M.~Freer, H.~Horiuchi, Y.~Kanada-En'yo, D.~Lee, U.G. Mei\ss{}ner, Rev. Mod.
  Phys. \textbf{90}, 035004 (2018).

\bibitem{C14-lin-chain.23}
J.~Han, Y.~Ye, J.~Lou, X.~Yang, Q.~Li, Z.~Yang, Y.~Yang, J.~Wang, J.~Xu, Y.~Ge
  et~al., Comm. Phys. \textbf{6}, 220 (2023).

\bibitem{Otsuka-clust.22}
T.~Otsuka, T.~Abe, T.~Yoshida, Y.~Tsunoda, N.~Shimizu, N.~Itagaki, Y.~Utsuno,
  J.~Vary, P.~Maris, H.~Ueno, Nat. Comm. \textbf{12}, 2234 (2022).

\bibitem{ER.04}
J.L. Egido, L.M. Robledo, Nucl.~ Phys. A \textbf{738}, 31 (2004).

\bibitem{ASPG.05}
P.~Arumugam, B.K. Sharma, S.K. Patra, R.K. Gupta, Phys. Rev. C \textbf{71},
  064308 (2005).

\bibitem{RMUO.11}
P.G. Reinhard, J.A. Maruhn, A.S. Umar, V.E. Oberacker, Phys. Rev. C
  \textbf{83}, 034312 (2011).

\bibitem{EKNV.14}
J.P. Ebran, E.~Khan, T.~Nik\ifmmode \check{s}\else
  \v{s}\fi{}i\ifmmode~\acute{c}\else \'{c}\fi{}, D.~Vretenar, Phys. Rev. C
  \textbf{90}, 054329 (2014).

\bibitem{YIM.14}
J.M. Yao, N.~Itagaki, J.~Meng, Phys. Rev. C \textbf{90}, 054307 (2014).

\bibitem{RA.16}
D.~Ray, A.V. Afanasjev, Phys.\ Rev. C \textbf{94}, 014310 (2016).

\bibitem{AA.18}
A.V. Afanasjev, H.~Abusara, Phys. Rev. C \textbf{97}, 024329 (2018).

\bibitem{ATT.24}
A.V. Afanasjev, S.~Teeti, A.~Taninah, Phys. Scripta \textbf{99}, 065313 (2024).

\bibitem{3s-density.PRL.82}
J.M. Cavedon, B.~Frois, D.~Goutte, M.~Huet, P.~Leconte, C.N. Papanicolas, X.H.
  Phan, S.K. Platchkov, S.~Williamson, W.~Boeglin et~al., Phys. Rev. Lett.
  \textbf{49}, 978 (1982).

\bibitem{3s-density-NPA.83}
B.~Frois, J.M. Cavedon, D.~Goutte, M.~Huet, P.~Leconte, C.N. Papanicolas, X.H.
  Phan, S.K. Platchkov, S.E. Williamson, Nucl. Phys. \textbf{396}, 409 (1983).

\bibitem{PA.22}
U.C. Perera, A.V. Afanasjev, Phys. Rev. C \textbf{106}, 024321 (2022).

\bibitem{PA.23}
U.C. Perera, A.V. Afanasjev, Phys. Rev. C \textbf{107}, 064321 (2023).

\bibitem{ZIM.15}
P.W. Zhao, N.~Itagaki, J.~Meng, Phys. Rev. Lett. \textbf{115}, 022501 (2015).

\bibitem{AARR.14}
S.E. Agbemava, A.V. Afanasjev, D.~Ray, P.~Ring, Phys.\ Rev. C \textbf{89},
  054320 (2014).

\bibitem{NL3*}
G.A. Lalazissis, S.~Karatzikos, R.~Fossion, D.P. Arteaga, A.V. Afanasjev,
  P.~Ring, Phys.\ Lett. \textbf{B671}, 36 (2009).

\bibitem{TO-rot}
A.V. Afanasjev, H.~Abusara, Phys.\ Rev. C \textbf{82}, 034329 (2010).

\bibitem{OFE.06}
W.~von Oertzen, M.~Freer, Y.~Kanada-En'yo, Phys.\ Rep. \textbf{432}, 43 (2006).

\bibitem{IMIO.11}
T.~Ichikawa, J.A. Maruhn, N.~Itagaki, S.~Ohkubo, Phys. Rev. Lett. \textbf{107},
  112501 (2011).

\bibitem{Ca40-PRL.01}
E.~Ideguchi, D.G. Sarantites, W.~Reviol, A.V. Afanasjev, M.~Devlin, C.~Baktash,
  R.V.F. Janssens, D.~Rudolph, A.~Axelsson, M.P. Carpenter et~al., Phys. Rev.
  Lett. \textbf{87}, 222501 (2001).

\bibitem{RS.80}
P.~Ring, P.~Schuck, {\em The Nuclear Many-Body Problem} (Springer-Verlag,
  Berlin)  (1980).

\bibitem{VDS.83}
M.J.A. de~Voigt, J.~Dudek, Z.~Szyma{\'n}ski, Rev.~ Mod.~ Phys. \textbf{55}, 949
  (1983).

\bibitem{SGBGV.89}
Y.R. Shimizu, J.D. Garrett, R.A. Broglia, M.~Gallardo, E.~Vigezzi, Rev.\ Mod.\
  Phys. \textbf{61}, 131 (1989).

\bibitem{KALR.10}
S.~Karatzikos, A.V. Afanasjev, G.A. Lalazissis, P.~Ring, Phys.\ Lett. B
  \textbf{689}, 72 (2010).

\end{thebibliography}

\end{document}